\documentclass[runningheads]{llncs}
\usepackage[T1]{fontenc}
\usepackage{graphicx}
\usepackage{soul}
\usepackage[T1,T2A]{fontenc}
\usepackage[utf8]{inputenc}
\usepackage{hyperref}
\usepackage{xcolor}

\begin{document}

\title{
Weaponizing Disinformation Against Critical Infrastructures
}

\author{Lorenzo Alvisi\inst{1,2}\orcidID{0009-0007-4222-348X}\and
John Bianchi\inst{1} \orcidID{0009-0006-2582-1480}\and
Sara Tibidò\inst{1,3}\orcidID{0009-0004-0646-0558} \and
Maria Vittoria Zucca\inst{1,4}\orcidID{0009-0004-0049-9611}
}
\authorrunning{L.Alvisi, J.Bianchi, S.Tibidò, and M.V.Zucca}
\institute{
IMT School for Advanced Studies, Lucca, Italy\\
\email{[name.surname]@imtlucca.it}\\
\and 
Institute of Informatics and Telematics, National Research Council (IIT-CNR), Pisa, Italy\and
University of Bari "Aldo Moro", Bari, Italy\and
Sant’Anna School of Advanced Studies, Pisa, Italy
}
\maketitle              

\begin{abstract}  

For nearly a decade, disinformation has dominated social debates, with its harmful impacts growing more evident. Episodes like the January 6 United States Capitol attack and the Rohingya genocide exemplify how this phenomenon has been weaponized. While considerable attention has been paid to its impact on societal discourse and minority persecution, there remains a gap in analyzing its role as a malicious hybrid tool targeting critical infrastructures. This article addresses this gap by presenting three case studies: a hypothetical scenario involving the electric grid, an attack on traffic management, and XZ Utils backdoor. Additionally, the study undertakes a criminological analysis to comprehend the criminal profiles driving such attacks, while also assessing their implications from a human rights perspective. The research findings highlight the necessity for comprehensive mitigation strategies encompassing technical solutions and crime prevention measures in order to safeguard critical infrastructures against these emerging threats.

\end{abstract}

\section{Introduction}
To fully grasp the notion of information fabrication, it's crucial to be well-versed in its origins and historical evolution. The "disinformation chronicles" trace back to ancient Rome, where intentional falsehoods were exemplified in the final years of the Republic (33 BC) through Octavian and Marcus Antonius’s disinformation campaigns against each other. Octavian’s successful use of targeted disinformation to ruin Antonius’ reputation ultimately solidified his power as Augustus, the first Roman Emperor. This maneuver marks a significant historical moment, showcasing how fabricated communication tactics can destabilize entire political systems\cite{van2024disinformation}. While disinformation’s roots are ancient, its impact has been amplified through history by the arrival of new means of communication. From the 15th-century printing press and the consequential rise of newspapers to the advent of radio, television, and photojournalism in the 20th century, until the later arrival of the Internet, followed by social media proliferation in the 21st century, prioritizing virality over veracity and creating unprecedented opportunities for (dis)information to reach its targets. Semantically, the term “disinformation” did not enter English dictionaries until the 1980s. It originated as a translation of the Russian дезинформация (transliterated as \textit{dezinformatsiya}), which traces its roots back to the 1920s when Russia employed it in connection with a special disinformation office whose purpose was to disseminate “false information as a strategic weapon with the intention to deceive public opinion”\cite{TheCyberDefenseReview}. \\ As a necessary methodological premise, the authors align with the  European Commission's definition of disinformation as stated in the 2022 Code of Practice on Disinformation\cite{EUCodeDisinfo}, namely as “verifiably false or misleading information” which, cumulatively, (a) “is created, presented and disseminated for economic gain or to intentionally deceive the public”; and (b) “may cause public harm”, intended as “threats to democratic political and policymaking processes as well as public goods such as the protection of EU citizens’ health, the environment or security”. This notion of dis-information encompasses both falseness and harmfulness, thereby distinguishing it from other forms of information disorder as outlined by the Council of Europe: misinformation, i.e., information that is false but not created with the intention of causing harm, and mal-information, i.e., genuine information used with the intent to inflict harm \cite{wardle2017information}.\newline
In this paper, we will focus on the impactful aspect of disinformation, examining it as a potential tool aimed at targeting and destabilizing national critical infrastructures, aligning closely with the “umbrella concept” of hybrid threats. Given the multifaceted nature of hybrid threats, which encompasses polymorphism, adaptability, undetectability, offensiveness, disruption, and manipulation, we will employ this framework to address the following question: \textit{how can disinformation be weaponized as a hybrid tool targeting vital strategic infrastructures?}
To this aim, the research adopts the following structure: Section \ref{sec:literature} reviews the current state-of-the-art and gaps in the literature regarding disinformation attacks on critical sectors. In Section \ref{sec:metodology}, we discuss three selected case studies, investigating the various facets of the relationship between disinformation and the security of critical infrastructures. Section \ref{sec:criminological}
provide a criminological overview of the potential actors behind the analyzed disinformation attacks, along with their criminal structures and motivations, while Section \ref{sec:impacts} assesses the implications of the investigated cases from a human rights perspective. Finally, Section \ref{sec:concluding} presents the conclusions.

\section{Literature Review}\label{sec:literature}

In order to clarify the core objective and substantial contribution of the research to the field, it is imperative, from the outset, to outline the present state-of-the-art concerning disinformation as a hybrid threat to critical infrastructures, therefore scrutinizing existing gaps in the literature that need to be addressed.\\
The main questions that were addressed in the past year have predominantly stemmed from real events that began online and ended up causing disruption in the real world. Events such as Brexit, PizzaGate, and the attack on Capitol Hill have underscored how disinformation can undermine the fundamental democratic principle of a State. 
Following media and prominent events, researchers have primarily focused on disinformation as a "political debate landscaping tool", leading studies concentrating on social media as the controlled environment where public debates occur and can be analytically measured. Consequently a broad series of studies on social networks has emerged\cite{GAMBINI202425} \cite{Avalle2024}, focusing on infodemics and echo chambers, conspirative communities \cite{alvisi2024unraveling}, and particular topics vulnerable to misinformation\cite{tardelli2024temporal}\cite{bianchi2024evaluating}, such as the covid-19 pandemics \cite{TardelliCalamusa2020}, climate change \cite{FalkenbergGaleazzi2022Climate} or elections \cite{Tardelli2024USelection}. The study of echo chambers is indeed particularly prolific since they play a crucial role in fostering polarization and radicalization \cite{cinelli2021echo}. This phenomenon has been associated with online toxicty\cite{Avalle2024}, but can also contribute to a spillover effect leading to offline violence and protests \cite{vasist2023polarizing}. Moreover, regarding critical sectors, the current state of knowledge appears well-explored within the governmental sphere, where a thorough examination of the ramifications of disinformation attacks on the electoral contexts exists. In fact, there are several recurring key narratives in election-related disinformation campaigns. These include false claims of widespread voter fraud and rigged processes, narratives aimed at voter suppression through misinformation about polling stations, and efforts to delegitimize election results by alleging fraud \cite{hybridcoe2024}. \\
However, the literature appears to be lacking regarding potential disinformation attacks targeting other critical infrastructures, which are equally strategic for providing national essential services and ensuring the well-being of society as a whole. Consequently, our research will proceed to present three case studies, focusing respectively on the electricity infrastructure \cite{CaseStudy1Raman}, the road system \cite{CaseStudy2Waniek2021}, and ultimately, the XZ backdoor. We have opted to concentrate on the first two case studies since they provide an innovative perspective on the impact of disinformation, which not only targets public debate but also demonstrates the potential for destructive consequences in undermining the daily lives of citizens. Indeed, to the best of our knowledge, few other studies have been conducted in this field, but among them, existing research highlights airline traffic \cite{Jamalzadeh2022} and railway networks \cite{JAMALZADEH2024109819} as vulnerable to disinformation.
Ultimately, we have selected the case study of the XZ backdoor for its demonstration of an innovative application of disinformation techniques, specifically from a social engineering
perspective, and its relatively unexplored status in current literature up to this point.

\section{Methodology}\label{sec:metodology}
This section analyzes the three case studies introduced in Section \ref{sec:literature}. For each case, we will examine the reasons that could lead to such an attack, the methods used or that could be used, and the consequences and reactions that have caused.

\subsection{Case Study 1 - Electic Grid }\label{sec:CS1}

Critical infrastructure security has been repeatedly threatened by disinformation. One of the most notable instances of this was at the onset of the COVID-19 pandemic, where disinformation triggered widespread panic, leading to mass stockpiling of essential goods \cite{micalizzi2021stockpiling}. However, this paragraph will not delve into this well-documented phenomenon but will instead focus on the scenario hypothesized by Raman et al. in the article \textit{"How weaponizing disinformation can bring down a city's power grid".} This article examines a hypothetical attack on a power plant to study the connection between disinformation and blackouts \cite{DBLP:journals/corr/abs-1908-02589}.

\subsubsection{Reason}\label{sec:powerReason}
Power plants are fundamental to the economy and citizens' security, providing essential energy for homes, businesses, and critical services such as hospitals, schools, transportation, and communication. It's no coincidence that in wartime scenarios, they often become points of interest \cite{griffith1994strategic}. Power plants are also one of the key pillars of a city's economy. In the study by Rose et al. \cite{rose2007business}, a total blackout lasting two weeks in Los Angeles County was simulated. The results of this simulation highlight the severe economic and social consequences of a prolonged power outage. According to the study, such an event would cause a loss of \$20.5 billion in terms of disrupted economic activities, demonstrating how deeply and pervasively our society depends on electrical energy.

\subsubsection{Methods}\label{sec:powerMethods}
The study's authors\cite{DBLP:journals/corr/abs-1908-02589} surveyed 5,124 participants through Amazon Mechanical Turk\footnote{\href{https://www.mturk.com/}{Amazon Mechanical Turk (MTurk)}}. Participants were presented with a message about a 50\% discount on electricity rates from 8:00 PM to 10:00 PM and were asked about the likelihood of changing their electricity consumption and sharing the message with others. The researchers examined two factors: the sender of the notification (a stranger or a friend) and the content of the notification (whether or not it included an external link). Based on these factors, participants were divided into four groups, and their responses were analyzed. While recognizing that survey behavior might not perfectly reflect real-life actions, the researchers used the responses to estimate participants' actual probabilities of follow-through. The results indicated that follow-through rates varied significantly depending on the network model and the value of k (the number of friends to whom a node forwards the message), ranging from 3.2\% to 26.8\%. The study found that messages without an external link had higher follow-through rates. Considering a context where 15\% of the population owns electric vehicles (EVs) and 30\% of the population is initially targeted by the disinformation, the resulting follow-through could cause blackouts affecting from 5.6\% to 100\% of residents, depending on the follow-through rate. The study concludes that behavioral manipulation through disinformation can potentially lead to significant disruptions in a heavily loaded power grid.

\subsubsection{Consequences and reactions}
The consequence of a massive blackout, due to our dependency on electricity, affects all kinds of essential services. One of the most impacted infrastructures is the healthcare system. As the blackout lasts, people, to cope with the moment, have an increase in consumption of alcohol and drugs\cite{jani2006hurricane} and spoiled food \cite{kosa2012older}, which could cause foodborne illnesses \cite{doi:10.2105/AJPH.2004.061358}. Also, mental health is affected as prolonged interruptions increase the likelihood of developing anxiety, depression, stress, and, in some cases, post-traumatic stress disorder (PTSD) \cite{gros2012relations}. In addition, the incorrect usage of generators can lead to increased hospitalization due to carbon monoxide poisoning \cite{van2007carbon}. Moreover, those who need regular medical treatment, such as dialysis\cite{abir2013impact}, or need special equipment\cite{miles2014socio} are at increased risk.
Another infrastructure that is easily and heavily impacted is the transport system. During the 27th March 2015 power outage in Holland, not only all the electricity-driven public transportation was disrupted for many hours, thus forcing people to travel by car, but also the traffic lights, creating chaos and congestion in the transport network. In this case, traffic speed decreased to 40$\%$ \cite{MELNIKOV2015336}. We also cannot overlook the economic impact as the Office of Technology Assessment of the U.S. Congress stated that the potential cost of a widespread power outage ranges from \$ $1/$kWH to \$ $5/$kWH of disrupted service. This value depends on the duration of the outage, the number of affected customers, and a variety of other factors. For example, the New York City outage of 1977 caused \$155 Million dollars of only arson and looting alone\cite{OTA1990}.

\subsection{Case Study 2 - Road Network}\label{sec:CS2}
The second case study, drawing from existing literature\cite{CaseStudy2Waniek2021}, involves simulating a disinformation attack aimed at disrupting the traffic network of the city of Chicago. While existing literature demonstrates the vulnerability of airplane\cite{Jamalzadeh2022}  and railroad traffic\cite{JAMALZADEH2024109819} to disinformation, we focus our attention on the road network due to its lower risk and cost for attackers, as it is less controlled and is a more disruptive target. We cannot overlook the fact that this could serve as either support or a decoy attack on another target, which may now be more vulnerable.

\subsubsection{Reason}\label{sec:TrafficReason} An attack aimed at increasing traffic congestion in a city could serve multiple purposes. Firstly, it could create chaos and elevate the sense of vulnerability and insecurity among the citizens. Secondly, it could inflict economic damage on a particular area, thereby impacting specific cultural or social classes of individuals \cite{Moyano2021trafficcongestion}. Thirdly, such an attack could distract law enforcement and emergency services, potentially facilitating further attacks on other, more vulnerable targets. Lastly, based on recent history, this type of attack could be politically motivated, as exemplified by the Bridgegate scandal in 2013.

\subsubsection{Methods}\label{sec:TrafficMethods} The reference cited employs a parameter to estimate how a disinformation campaign, done by text, will be followed and its subsequent impact on traffic journeys. For our scenario, instead of estimating how many people would follow a disinformation attack, we will analyze an adversarial attack on a traffic map provider, as was done by Simon Weckert\footnote{\href{https://www.youtube.com/watch?v=k5eL_al_m7Q\&t=4s\&ab_channel=SimonWeckert}{Simon Weckert Youtube Channel}}. This approach allows us to estimate how many people would follow the direction suggested by said provider instead of the disinformation campaign. This method is clearly safer and less costly for the attacker. Given our scenario, we will consider just the case in which the attacker can create a convergence attack and cannot choose random points to avoid.

\subsubsection{Consequences and reactions}\label{sec:TrafficConsequences}
To analyze the effectiveness of the adversarial attack on the GPS traffic data we will reference studies highlighting the dependency on GPS apps. The Pew Research Center\footnote{\href{https://www.pewresearch.org/internet/2015/04/01/us-smartphone-use-in-2015/}{The Pew Research Center - Smartphone use in 2015}} stated that, already in 2015, 31$\%$ of drivers frequently used their navigation apps for turn-by-turn navigation, and given the increase in the market share of navigation apps\footnote{\href{https://www.statista.com/outlook/amo/app/navigation/united-states}{Data from Statista.com}} since then, we can safely assume that that number has drastically increased. Utires\footnote{\href{https://www.utires.com/articles/where-drivers-need-gps-the-most/}{Data from utires.com}} found out that $17\%$ of Americans could not reach their destination without using a GPS application and that half of the country relies on GPS for navigation. It was also noted \cite{laor2022waze} that WAZE users (a driving app) exhibit behavioral patterns that are in line with the four symptoms of technological addiction, thus reinforcing the idea that drivers will follow the app instructions without thinking if the instructions they receive are correct.

While we do not seek a precise estimate of how many drivers will blindly follow the driving app of their choice, based on the aforementioned data, it's safe to assume that it will be higher than 15$\%$. This assumption places us in the worst-case scenario outlined in our reference paper. As a result, around the target area, there will not be roads with decreased traffic, and the vast majority of the roads close to the target area will experience a significant increase in vehicle numbers, creating congestion in the target's proximity. This result remains consistent with some previous literature that noted how \cite{bianchin2020routing} routing apps may deteriorate stability in traffic networks.

\subsection{Case Study 3 - XZ Backdoor}\label{sec:CS3}
This section discusses the cyber attack on XZ Utils\footnote{\href{https://zlib.net/}{XZ Utils - Official website}} in which the attackers demonstrated a sophisticated strategy by compromising the software update system. XZ Utils, primarily maintained by Lasse Collin, was found vulnerable as Collin was weighed down by mental health issues and temporarily abandoned the project. During this period, a newcomer named 'Jia Tan' emerged, swiftly gaining Collin's trust and assuming responsibility for updates. However, unbeknownst to Collin, Jia Tan used this position to push packages that would, in the future, allow the operation of the backdoor, thereby indirectly compromising the system. However, it is unclear whether Jia Tan was aware of the backdoor's presence in the code. 
The incident underscores the dependence on vulnerable technologies and open-source software, which are often managed by unsupported volunteers. There is suspicion of state actor involvement, similar to the SolarWinds attack \cite{alkhadra2021solar}, prompting reflection on two fundamental aspects: the fragility of technological foundations and the critical role of often overlooked open-source maintainers \cite{naughtonGuardian}.

\subsubsection{Reason}\label{sec:XZReason}

Before delving into the attackers' methodologies step by step, it's important to understand why this particular software was chosen, seeking to grasp the requirements they pursued. XZ Utils is an open-source software \cite{zlib_manual} maintained by volunteers who manage the project in a hobbyist manner \cite{RooseNYTimes}. Widely utilized, it's a component of Linux-based operating systems  \footnote{\href{https://www.kernel.org/}{XZ data compression in Linux - The Linux Kernel Archives.}}, which are employed worldwide by 1.5\% of desktop systems and 62.7\% of servers \footnote{\href{https://www.fortunebusinessinsights.com/server-operating-system-market-106601}{Fortune Business Insights}}. The attackers' attention has focused on widely used server-side software. XZ Utils emerges as the perfect target for this type of requirement and is vulnerable to a supply chain attack via social engineering.

\subsubsection{Methods}\label{sec:XZMethods}
In October 2021, an individual using the pseudonym "Jia Tan" began contributing to the XZ Utils project via the \textit{xz-devel mailing list}\cite{XZMailingList}. By December 2022, Tan was allowed to add code directly to the project with community approval\footnote{\href{https://github.com/tukaani-project/xz/commit/8ace358d65059152d9a1f43f4770170d29d35754}{JiaT75, "CMake: Update .gitignore for CMake artifacts from in-source build," Commit on GitHub}}. In March 2023, Tan gained control over the \textit{"OSS-Fuzz"} test component\cite{serebryany2017ossfuzz}, effectively becoming a co-maintainer. Later, Hans Jansen (another user not traceable to an individual) submitted modifications using IFUNC\cite{gnu_ifunc}, which Tan approved. To hide the malicious code, Tan disabled IFUNC in OSS-Fuzz tests. On February 23, 2024, Tan added backdoor-containing test files to the project \footnote{\href{https://git.tukaani.org/?p=xz.git;a=commitdiff;h=cf44e4b7f5dfdbf8c78aef377c10f71e274f63c0}{git.tukaani.org - Jia Tan merges hidden backdoor binary code}}. The next day, version 5.6.0, which contained the backdoor, was released and subsequently incorporated into Linux-based distributions\footnote{\href{https://research.swtch.com/xz-timeline}{Timeline of the xz open source attack
}}\cite{Lins2024}.

\subsubsection{Consequences and reactions}\label{sec:XZConsequences}
On March 29, 2024, Red Hat and America's Cyber Defense Agency alerted all Fedora Linux 40 and Fedora Rawhide users about the presence of a backdoor in their systems. Due to its widespread presence on servers and the lack of limitations, the backdoor received a score of 10 out of 10 in the Common Vulnerability Scoring System\footnote{\href{https://access.redhat.com/security/cve/CVE-2024-3094}{RedHat Official Website}}. The backdoor was discovered thanks to Andres Freund, a programmer who noticed that SSH access was using a small amount of CPU, which prompted him to perform a check that is rarely done. This backdoor could execute any command as a superuser if the SSH protocol received a particular key followed by the commands instead of the user signature needed to complete the SSH connection. This complete lack of limitations on the attacker's capabilities makes it impossible to determine how many devices were breached, as they had the possibility to delete all logs containing their traces. This attack highlights two significant vulnerabilities within the global information infrastructure: first, the reliance of private companies on the open-source community, and second, the susceptibility of this community to social engineering and disinformation attacks.

\section{Criminological Analysis}\label{sec:criminological}
After thoroughly examining the diverse types of disinformation attacks via case studies, it is crucial at this stage of the research, to undertake an analysis of the malicious actors behind these assaults, namely those responsible for crafting and disseminating the disinformation vehicle. Specifically, we aim to address the following criminological questions: \textit{How can these malevolent agents be classified? Do they operate within a structured criminal organization? What motivations propel their actions?\\}
To this end, provided below is a general taxonomy of potential criminal profiles relevant to our discussion:\\
\textit{State-sponsored actors}: 
Undoubtedly, the last decade has seen a significant rise in state-sponsored cybercriminal activities, with threat actors entrenched within military or government agencies being particularly concerning due to their extensive resources, expertise, organization, and sophisticated methods \cite{ElectionLawJournal}. Among their specific offensive operations (e.g., cyber espionage, political disinformation campaigns), an emphasis is placed on targeting critical infrastructure \cite{Lehto2022}. Therefore, the analyzed disinformation attacks (Section \ref{sec:metodology} paragraph \ref{sec:CS2}, paragraph \ref{sec:CS1}) seem to align with these types of destructive operations, presenting a significant threat in a hypothetical scenario of hostile cyber warfare, owing to their potentially significant impact on national security and citizen safety. 
While the XZ case (Section \ref{sec:metodology}, paragraph \ref{sec:CS3}) holds significant relevance, as it carries the potential to escalate into cyber espionage, a common objective among contemporary state-sponsored groups. These attacks commonly target large corporations and government entities, aiming to illicitly access their systems to gather intelligence on critical sectors, ultimately enhancing their nation's security, economic competitiveness, and military capabilities \cite{rubenstein2014nation}. \\
\textit{Cyber-terrorists:} It is widely acknowledged that ICT can be utilized to promote, support, facilitate, and/or engage in acts of terrorism. Although there isn't a universally accepted definition, the term "cyberterrorism" has gained traction in literature, describing it as a cyber-dependent crime committed for ideological purposes, aimed at instilling fear, intimidation, and/or coercion within a targeted government or population, with the intent to cause or threaten harm \cite{jarvis2014cyberterrorism}. However, while ideology provides a broad rationale for the targeting of terrorist groups, the selection of specific targets within the spectrum of ideologically acceptable ones is influenced by other factors, which are best described as strategic or tactical \cite{ackerman2006assessing}.For instance, an essential factor influencing terrorist targeting is the level of protection of a facility, as (cyber)terrorists would be more inclined to choose targets that are vulnerable. Simultaneously, they aim to attack functionally crucial, high-profile targets whose destruction would inflict significant costs on the host society. Critical infrastructures hold particular significance in this assessment, given that an attack against them could severely compromise national security, economic stability, and social welfare. In this regard, it is noteworthy to mention the case study analyzed concerning the disinformation attack on the electricity power grid (Section \ref{sec:metodology}, paragraph \ref{sec:CS1}), as it aligns with a conceivable cyberterrorist scenario, potentially leading to massive blackouts or even temporary regional power disruption, widespread public fear, and an image of helplessness that would directly serve the terrorists' objectives \cite{national2012terrorism}. \\
\textit{Hactivists:} While a universally agreed-upon definition of hacktivism is yet to be established, it has been described as the fusion of “hack” and “activism”, denoting the nonviolent utilization of illegal or legally ambiguous digital tools to effect social or political changes \cite{samuel2004hacktivism}. The hacktivist landscape, less structured than cybercriminal activities, comprises individuals and groups with diverse skills and capabilities who may act independently as "lone wolves" or collaborate within decentralized, transnational collectives, forming temporary groups for specific orchestrated operations. Since these actors carry out political activism leveraging the Internet to create an impact and exploit security vulnerabilities to achieve their objectives, the feasibility of their operations targeting critical infrastructure seems plausible. In particular, hacktivist behaviors may involve intentionally accessing systems, websites, and/or data without authorization, intentionally interfering with the functioning and/or accessibility of systems, and stealing and exposing sensitive information. As evidenced in cases like XZ  (see in section \ref{sec:metodology}, paragraph \ref{sec:CS3}), disinformation campaigns, including the use of fake accounts, can facilitate such intrusions, ultimately aimed at stealing sensitive information. In the case of hacktivists, their aim may be to embarrass the organization by highlighting the laxness of its information security rather than solely obtaining the information itself. Subsequently, they may choose to release this information publicly or exploit it to further their political or social objectives \cite{citation-key}. \\
\textit{Cybercriminals:} It's well-established that digital technologies have “democratized” crime, enabling even small or solitary actors to execute complex malicious tasks and criminal schemes \cite{wall2024cybercrime}. However, it's important to note that cybercrime, as an "umbrella term", covers a broad spectrum of distinct criminal activities, with our focus herein directed towards cyber-dependent illicit activities facilitated and targeted through the use of ICTs \cite{mcguire2013cyber}. Traditionally limited to highly skilled individuals working independently, cybercriminal activities have evolved to include sophisticated organizations recruiting technical experts to oversee networks of "businesses," resulting in significant financial profits. An example is the “Crime-as-a-Service” business model, through which a smaller affiliated criminal group will rent the “ready-to-use” malicious package from a larger cyber group to execute their attacks effectively.  Hence, it is therefore plausible that within the web underground (and even within the surface web), marketplaces offering Disinformation-as-a-Service (DaaS) may thrive. Here, potential offenders can acquire various resources, such as fake accounts (for their disruptive utilization, reference can be made to case XZ), AI-generated multimedia content, and all the requisite tools to orchestrate sophisticated disinformation campaigns \cite{citationnyt}.\\

\section{ Human Rights Implications }\label{sec:impacts}

The European Union, in the Council Directive 2008/114/EC on "the identification and designation of European critical infrastructures and the assessment of the need to improve their protection", defines critical infrastructures as "an asset, system or part thereof (...) which is essential for the maintenance of vital societal functions, health, safety, security, economic or social well-being of people, and the disruption or destruction of which would have a significant impact on a Member State due to the failure to maintain those functions" \cite{CouncilDirective}. Therefore, while critical infrastructures are crucial for the well-being and stability of a society, they are also intrinsically linked to fundamental rights and freedoms. Indeed, these infrastructures encompass services and facilities necessary to ensure a minimum standard of living, and any degradation or interruption in their supply would significantly impact the safety and security of the population and the functioning of state institutions.\\
However, the interdependency that exists between cyber-physical-social networks makes them more vulnerable to large-scale disruption \cite{barker2017defining}.  Consequently, a malfunction in one of these structures can easily propagate to others, magnifying its effects and triggering a domino effect of violations, including those concerning fundamental human rights. States have the duty, under international human rights law, to safeguard the human rights of individuals within their territory and/or jurisdiction, even from third parties' abuse or interference \footnote{According to, for example, Article 2 of the "International Covenant on Civil and Political Rights" and the "International Covenant on Economic, Social and Cultural Rights", and on what set on the "Guiding Principles on Business and Human Rights"}. This obligation is particularly important considering the potential impacts that attacks on critical infrastructures may have on individuals and communities \cite{UNRCCA}. To illustrate the potential domino effect of cyberattacks on these infrastructures, which could lead to adverse impacts on human rights, reference can be made to scenarios outlined in the case studies presented in Section \ref{sec:metodology} of this paper.
Art.12 of the "International Covenant on Economic, Social and Cultural Rights" (ICESCR) states that States Parties "recognize the right of everyone to the enjoyment of the highest attainable standard of physical and mental health", and that the steps undertaken to achieve the full realization of the right include "the creation of conditions which would assure to all medical service and medical attention in the event of sickness" \cite{ICESCR}. Indeed, the right to health is an inclusive right that also encompasses the elements of accessibility and availability: health services should be timely, by reducing waiting times and delays, and should provide a sufficient quantity of functioning health facilities, goods, and services \cite{WHO}. Therefore, a hypothetical attack on a power plant may affect this right since it is possible, for example, that most doctors' surgeries or medical centers do not have any emergency power capabilities. While general medical practices can maintain rudimentary operations without using mains-dependant equipment, specialist doctors' surgeries, relying on specialist technology, cannot operate without electricity \cite{Blackout}.\\
Similarly, an attack on traffic networks can increase traffic congestion, thus making it difficult or requiring excessive time to reach a place. This can undermine the accessibility of people to health services since ambulances can have difficulties reaching a place or cars reaching hospitals. Traffic congestion thus reduces the performance of first responders, creating delays in fire trucks responding to fires, providing emergency medical services, and responding to other emergencies \cite{BRENT2020102339}. Moreover, traffic congestion, according to the scope and duration of the situation, may also risk impacting the supply chain of basic necessities, at least in the affected area, thus potentially undermining the right to an adequate standard of living as enshrined in Article 11 ICESCR \cite{ICESCR}.\\
At the time of writing, the authors of this paper are not aware of cases in which damages can be led back to the XZ backdoor, nor are they sure it will actually be possible to link damage to it safely. Nevertheless, it can be supposed that at high risk may be the right to privacy, safeguarded by Article 8 of the "European Convention of Human Rights"(ECHR) \cite{ECHR} and Article 17 ICCPR \cite{ICCPR}. Indeed, it is conceivable that an attack of this nature could interfere with the computer systems of critical infrastructures that house and process vast amounts of sensitive and private information. Such an intrusion, coupled with potential exfiltration or alteration of data, would constitute a breach of the right to privacy and personal data protection and would contribute to reputational and economic damage.\\
The European Court of Human Rights, in the case Podchasov v. Russia, recognized the violation of the right to private life under Article 8 ECHR when, in July 2012, the Russian Federal Security Service (FSB) required "Telegram Messanger LLP" to disclose technical information in order to facilitate the decryption of communications between some Telegram users suspected of "terrorism-related activities" \cite{EUCHRj}. However, Telegram refused to comply, stating that it was "technically impossible" to execute FSB's order "without creating a backdoor that would weaken the encryption mechanism for all users" \cite{EUCHRj} and thus violating their right to privacy. The XZ case highlights a different situation, where the backdoor was not created to facilitate investigations. However, the scope and concerns remain the same for everyone affected by the XZ attack, in Europe and outside. In the same judgment, the European Court reported a joint statement by Europol and the European Union Agency for Cybersecurity (ENISA) affirming that introducing backdoors "while this would give investigators lawful access in the event of serious crimes or terrorist threats, it would also increase the attack surface for malicious abuse, which, consequently, would have much wider implications for society" \cite{EUCHRj}. Therefore, it is possible to foresee that the extent of the consequences of an attack such as the one conducted in XZ, although difficult to trace, can be extensive and severe, also impacting other rights and leading to economic damage. For example, intellectual property rights, which potential acts of cyber espionage could violate, can cause damage to a company in terms of the cost of cleaning up the systems that have been attacked, opportunity cost, negative impacts on innovation, and reputational damage \cite{EUimpact}.

\section{Concluding Remarks}\label{sec:concluding}
In conclusion, it is clear from the above discussion that disinformation can serve as a (hybrid) tool for launching effective attacks, resulting in detrimental impacts on the physical security of critical infrastructures. This highlights the importance of prioritizing future research efforts to address and bridge the existing gap in the literature. However, several open questions remain on the topic of prevention and countermeasures strategies, as briefly outlined below. \\
Firstly, from a traditional law enforcement perspective, various factors contribute to the challenges faced in investigating, tracing, and countering these types of cybercrimes, including: i) ineffectiveness in tracing criminal activity; ii) difficulty in attributing ownership and authorship; iii) law enforcement officers and prosecutors lacking the technical expertise needed to handle cybercrime cases; iv) police lacking specialized tools for extracting information or sufficient computational power to process data expeditiously; v) strict and formal international cooperation mechanisms; vi) legislative provisions that are not harmonized among members of the international community, which lead to difficulties in legally classifying these malicious actions under national criminal law\cite{brown2015investigating}.  \\
Secondly, there is a lack of clarity regarding the mitigation strategies, including technical, security, and organizational measures, that should be implemented to align with the current EU regulatory framework for protecting critical infrastructures against emerging cyber threats \cite{EU-2557}. The following are some key considerations on this topic. As an initial reflection, it is worth noting that assessing the risk of such attacks is intrinsically challenging because disinformation serves as the vector of the attack, not the final goal. In these scenarios, attackers spread false information with the intention of manipulating societal behavior. However, the actual damage is often unknowingly inflicted by ordinary citizens who believe and act upon the false information they encounter. This makes it extremely difficult to trace the origins of the disinformation and hold the real perpetrators accountable. Furthermore, it raises ethical concerns about how to respond. Persecuting individuals who have been misled into spreading false information themselves can be seen as morally (and legally) questionable, as they are victims of the deception rather than the perpetrators. This complexity underscores the actual need for sophisticated strategies to both prevent the spread of disinformation and mitigate its impacts without unjustly targeting innocent citizens.\\
Given our present lack of readiness and vulnerability to such attacks, it is crucial to formulate a response strategy. This strategy can be divided into two parts: proactive measures and reactive responses. Proactively, we aim to minimize the effects of disinformation. Reactively, we must promptly stop the attack as soon as it begins.
Building on the strategies crafted in the fight against disinformation, we hope to find effective methods to respond to these attacks. Indeed, the progress made in enhancing public awareness and assisting individuals in distinguishing misinformation from truth has been significant. Given these advancements in collective awareness, we can leverage this progress to conduct campaigns to reduce the influence of disinformation attacks, especially those potentially tied to terrorist motives, thus safeguarding national security.\\
For the reactive part, our strategy should be built on two pillars: monitoring and reporting. Monitoring involves the continuous and systematic observation of critical infrastructures to detect any anomalies or unusual patterns that could indicate the presence of an attack. This includes tracking metrics such as traffic flow on major roadways and electricity usage across the grid. These indicators are crucial because significant, unexplained spikes are rare and typically have identifiable causes, ranging from natural events to technical failures or malicious activities. By maintaining a vigilant watch over these metrics, we can promptly spot potential issues as they arise.
Since we are discussing monitoring, it is crucial to emphasize that our focus is directed towards the entire system rather than individual citizens. This means monitoring traffic flow or electricity usage as a whole rather than the specific routes taken by individuals or their personal energy consumption. By doing so, we can effectively safeguard privacy and personal freedoms, avoiding issues related to invasion of the personal sphere or infringement on individual liberties.



\section{Contributions}
The authors' names order is alphabetical. All authors contributed jointly to the design of the research and the structure of the paper and jointly wrote Sections 1, 2, and 6. L.A. and J.B. wrote Section 3, M.V.Z. wrote Section 4, and S.T. wrote Section 5.
\bibliography{main}
\end{document}